\documentstyle[12pt]{article}

\def\ep{\epsilon^\prime}
\def\n{\nu}
\def\l{\lambda}

\def\l{\lambda}

\newcommand{\be}{\begin{equation}}
\newcommand{\ee}{\end{equation}}
\newcommand{\bea}{\begin{eqnarray}}
\newcommand{\eea}{\end{eqnarray}}

\begin{document}
\section*{An apprach to generate large and small leptonic mixing angles} 

\large{Biswajoy Brahmachari}

\vskip .5cm

\noindent {Department of Physics and Astronomy\\
University of Maryland\\
College Park, MD-20742}

\begin{abstract}
We take up the point of view that Yukawa couplings can be either 0 or 1, 
and the mass patterns of fermions are generated purely from the structure of 
the Yukawa matrices. We utilize such neutrino as well as charged leptonic 
textures which lead to (maximal) mixing angles of $\pi/4$ in each 
sector for relevant transitions. The combined leptonic CKM mixing 
angles are $\pi/4 \pm \pi/4$ which lead to very small $\sin^2 2 \Theta$ 
relevant to solar neutrino and LSND experiments. We propose that on the 
other hand the absence of the charged leptonic partner of the sterile 
neutrino maintains the angle $\pi/4$ from the neutrino sector for the 
transition $\nu_\mu \leftrightarrow \nu_s$ and hence atmospheric neutrino 
anomaly is explained through maximal mixing.

\end{abstract}

\noindent Recent `evidence' of neutrino mass detected at Super Kamiokande 
experiments\cite{supkam} and announced at the neutrino-98 
conference\cite{announcement} has sparred new enthusiasm for the studies of 
physics beyond standard model, particularly of the neutrino mass matrices
in the leptonic sector. In the Standard model (SM) neutrinos are massless 
and being based on gauge symmetries SM treat all the generations of matter 
identically though it is well-known that fermion masses differentiate among 
generations. If neutrinos have masses they are not expected to be 
generation blind at the same token. Neutrino oscillations establish 
this. The neutrinos should not only be massive, but also different 
generations must differ in masses for them to oscillate into one another. 
At a deeper level (in a gauge theory) we relate the masses to the Yukawa 
matrices or textures describing the interactions of fermionic matter with 
scalars whereas their forms are left unconstrained by gauge symmetries. 
Even-though grand unified gauge symmetries may relate\footnote{This can 
be avoided if the masses of the quarks and leptons arize from different 
scalars.} the Yukawa interactions of the quarks to those of 
leptons\cite{lee,lee1} the strengths of the interactions still remain 
arbitrary. Thus, it is imperative to study forms of textures as physics 
beyond standard model. Hence, we study simple symmetry properties of 
leptonic Yukawa textures themselves in the generation space which may 
describe the patterns of masses (eigenvalues) and mixing angles 
(eigenvectors) suggested by a variety of experimental measurements. 
For example, experimental inputs are provided by laboratory experiments as 
LSND neutrino oscillations experiments and neutrino-less double beta 
decay experiments, terestrial experiments such solar and atmospheric 
neutrino experiments as well as astrophysical observations such as 
nonluminous `dark matter' and the abundance of $^4$He in metal-poor blue 
compact galaxies.

Even if the preliminary data from Karmen experiments\cite{karmen} have 
failed to reproduce the LSND results; we $assume$ that in course of time 
LSND observations will be well established which indicates that 
$\overline{\nu_\mu}(\nu_\mu)$ is oscilatting to $\overline{\nu_e}(\nu_e)$ 
with $\Delta m^2_{e\mu}$ in the electron volt range. In this case solar 
neutrino deficit can be caused by $\nu_e \leftrightarrow \nu_\tau$ or 
$\nu_e \leftrightarrow \nu_s$ oscillations with $\Delta m^2$ 
approximately $10^{-5}$ eV$^2$; whereas the atmospheric neutrino anomaly 
can be explained by $\nu_\mu \leftrightarrow \nu_\tau$ 
or $\nu_\mu \leftrightarrow \nu_s$ oscillations, with $\Delta m^2$ 
approximately $10^{-3}$ eV$^2$. The pattern which emerges as a
result of the three experiments is that there must exist at-least four 
neutrino species\footnote{Three mass differences require at-least four 
neutrinos.} which are pair-wise degenerate in mass at the leading 
order and the pairs themselves are separated by a mass difference in the 
eV range. Symmetries which force a mass degeneracy among the pairs 
$\{\nu_e,\nu_\tau\}$ and $\{\nu_\mu,\nu_s\}$ has been recently discussed 
in reference\cite{anti}. Furthermore there is a need to understand why 
the angle relevant to the solar and LSND transitions are small whereas 
that of the atmospheric transition is maximal? In other words, from the 
point of view of model building, is there a natural way to deduce small 
angle MSW solution of the solar neutrino problem, small angle solution 
for the LSND observations and the maximal mixing to explain atmospheric 
neutrino anomaly? In this paper we address this question. Note that the mixing 
pattern is not solely the property of the neutrino mass matrices, as the 
neutrino mixing has to be folded in with the mixing of the charged 
leptons to get the realistic mould of leptonic mixing. There 
can be enhancement or cancellation of the mixing angles 
conceived in the neutrino sector due to `interference' with the 
charged leptonic mixing. We will use this interference of mixing angles
to isolate the large and small leptonic CKM mixing patterns. 
To do this we will have to supplement the $4 \times 4$ neutrino Yukawa 
textures with the $3\times3$ Yukawa textures of the charged leptons (which 
generates hierarchical pattern of charged lepton masses) such that the 
combined mixing matrix emerges in congruence with the pattern we are 
seeking. 

There are important astrophysical inputs towards the construction of 
the neutrino textures. $\Omega=1$ with $h \approx 0.5$ Cold Dark Matter 
cosmological models provide a good fit to the observational data in the 
presence of massive neutrinos, when $m_\nu \approx 5 eV$ is equally shared 
between two relatively heavy neutrinos, contributing a tiny Hot component 
(CHDM) to the dark matter\cite{primack}. As we will see below, our 
textures will not achieve this as that will give too much Majorana mass to 
the electron neutrino contradicting the bounds from the non 
observation of neutrinoless double beta decay described in the next 
paragraph. At best these textures 
can give a pair of neutrinos approximately at 1 eV. On the contrary, 
neutrinos in this mass range might modify the power spectrum to agree 
better with the data on galaxy distribution in the $\Omega \approx 0.4$, 
$\Omega_\Lambda \approx 0.6$ cosmology indicated by the high-red-shift 
supernova data\cite{primack1}, which may resolve the problems in r-process 
nucleosynthesis\cite{fuller} in Type II supernovae. Secondly 
there exists bounds on the product $\Delta m^2_{is}~\sin^4 2 \Theta_{is}$ 
form Big bang nucleosynthesis. These bounds are derived by demanding 
that oscillations do not bring the sterile neutrino in equilibrium 
with the known neutrino species. Moreover, if the number of electron 
neutrinos is depleted by oscillations to other species during the 
BBN epoch, the freeze-out temperature of the neutron-proton 
transition will also be increased. However, the bound derived from 
these depend on the 
value of the primordial lepton asymmetry. If the primordial asymmetry is 
of the order of $10^{-9}$ and the mixing is maximal, the bound upper 
bound from neucleosynthesis is nearly $\delta~m^2 \approx 
10^{-8}\cite{bbn1}$. However, if the initial lepton asymmetry is 
large enough $L_e \approx 10^{-5}$ the bound on the mixing between an 
ordinary and sterile neutrino is weakened and large-angle-mixing solution 
of the atmospheric neutrino anomaly becomes feasible\cite{bbn2}. 
Finally, the abundance of $^4$He also restricts the total number of 
neutrino 
species. The mass fraction of $^4$He termed $Y_P$ is obtained from the 
observation of metal-poor blue compact galaxies by linear extrapolation 
to zero nitrogen/oxygen abundance\cite{galaxy}. It was emphasized\cite{he4} 
that extra neutrino species would boost the 
reletivistic energy density during the big bang neucleosynthesis which 
increases the yield of Helium. The primordial mass fraction 
increases as $Y_P\approx 0.012 ~\Delta N_\nu$. However at the same time 
it also increases logerithmically with nucleon density $\eta\equiv 
n_N/n_\gamma$ which is somewhat uncertain. Taking $\eta\approx 2-9 
~~10^{-10}$, Karmen and Sarkar\cite{subir}has quoted that up to 1.5 
additional neutrino species are allowed consistent with Helium production 
at the neucleosynthesis era. Thus our scenario of one extra sterile 
species is in consonance also with astrophysical measurements.

Neutrinoless double beta decay is unobserved in nature. This leads to
a major constraint on the diagonal entry (1,1) of the neutrino 
Majorana mass matrix in the gauge basis. Unless the mixing matrix which 
rotates the mass basis to the gauge basis is fine tuned to cancel the 
(1,1) entry in the gauge basis, absence of double beta decay can give 
upper limits on the actual mass of the electron neutrino. The 
Heidelberg-Moscow experiments quote\cite{klapdor} the lower limits on the 
half life as $\Gamma_{1/2} \ge 1.1~10^{25}$ y; which already restricts 
the $<\nu_e\nu_e>$ Majorana mass term to be less that 0.60 eV at 90 $\%$ 
confidence level\footnote{The limit depends on nuclear matrix elements. 
See Table (4) of Ref\cite{klapdor}.}. In future the limits may go down to 
0.1 eV with the present experimental setup. We shall see that especially 
in combination with LSND results, which require $\Delta m^2_{e \mu} 
\approx 0.3$ eV$^2$ or higher (lower bound), neutrino-less double beta decay 
constraints (upper bound) has potential to rule out our textures of the 
neutrino mass matrices. 

Unlike the mass pattern of the neutrinos we are considering, the 
charged lepton masses are definitely hierarchical. We will stick with
three charged leptons with masses $m_e=0.51\times 10^{-3},~m_{\mu}=105.65 
\times 10^{-3},~m_\tau=1.777$ in GeV units. Mass matrices with 
hierarchical eigenvalues are well studied in the literature. We will take 
the point of view that the Yukawa couplings can be 0 or 1, meaning the 
textures can be filled with either 0 or 1. In such a scenario the mass 
pattern is dictated purely from the structure of the texture. This is 
very possible for the neutrino sector, neutrinos being gauge singlets 
their Yukawa couplings do not renormalize below the GUT scale. The 
charged lepton Yukawa couplings, however do renormalize from $SU(2)_L 
\times U(1)_Y$ interactions. However, the leptons being $SU(3)_{color}$ 
singlets avoid the major renormalization effects below the GUT scale 
which is due to QCD. If at the GUT scale we assume $h^2 / { 4 \pi}=1$, at 
the low energy the leptonic Yukawa couplings approach a quasi-infrared 
fixed point which is approximately $h \sim 1$, on the other hand 0 being 
a trivial fixed point of the renormalization group equations remain 
unaltered. An example of such a texture for quarks and leptons are given 
in reference\cite{fritzch}, where in the favor basis all the entries of 
the $3 \times 3$ Yukawa matrix are 1. This rank one matrix has two 
vanishing eigenvalues, thus giving a mass gap between 
the third generation (which is massive) and the lighter generations which 
have vanishing masses at the symmetry limit. The lighter generations 
acquire mass due to small breaking of the democratic symmetry. We will take 
this texture for the charged leptons in this paper which can be expressed as,
\be
M_l=c_l\pmatrix{ 1 & 1 & 1 \cr 
                 1 & 1 & 1 \cr
                 1 & 1 & 1} + 
\pmatrix{ \delta_1 & 0 & 0 \cr
          0 & -\delta_1 & 0 \cr
          0 & 0 & \delta_3} \label{a}
\ee
The parameters $\delta_i$ measure the democratic symmetry breaking, and 
they can be fitted with the known masses of the charged leptons. The 
mixing matrix of the charged leptonic sector in the symmetry limits 
given by, 
\be
O^\prime_L=\pmatrix{1/\sqrt{2} & -1/\sqrt{2} &0 &0\cr
      1/\sqrt{6} & 1/\sqrt{6} &-2/\sqrt{6}&0 \cr
      1/\sqrt{3} & 1/\sqrt{3} & 1/\sqrt{3}&0\cr
       0&0&0&1}, \label{onu1}
\ee
The correction to Eqn \ref{onu1} due the 
symmetry breaking can be expressed as, 
\bea
&& O^{\prime \prime}_L=\sqrt{m_e \over m_\mu}\pmatrix{ 1/\sqrt{6} & 
1/\sqrt{6} & -2 /\sqrt{6} &0\cr 
-1/\sqrt{2} & 1/ \sqrt{2} & 0 &0\cr
0 & 0 & 0 &0 \cr
0 & 0 &0 & \gamma_1} \nonumber \\
&& + {m_\mu \over m_\tau}\pmatrix{0 & 0 & 0 &0\cr
                            -1/\sqrt{6} & -1/\sqrt{6} & 1/\sqrt{6} &0\cr
                            1/2\sqrt{3} & 1/2\sqrt{3} & -1/\sqrt{3} &0 \cr
                             0 & 0 & 0 & \gamma_2}
\eea
In both the $O^\prime$ and $O^{\prime \prime}$ the rows represent the
eigenvectors of the matrix given in Eqn. (\ref{a}); and $\sqrt{m_e/m_\mu} 
\gamma_1 + m_\mu/m_\tau \gamma_2=0$. The top most being electron, the middle 
second is the muon
and the third is tau lepton. The realistic mixing in the leptonic sector 
after taking into account the corrections is 
\be
O_L=O^\prime_L + O^{\prime \prime}_L. \label{ol}
\ee
Note we have added a unit vector in the fourth row and fourth columns of 
the lepton mixing matrix. This can be thought of as due to the existence of 
a fictitious heavy lepton at the Plank scale which does not mix with 
anything and has no role to play at the low energy physics. We have 
introduced it as a trick to do our calculations of matrix multiplication 
as will be clear in a moment.

Now we briefly review the texture of the neutrino sector. Let us consider 
that an $n$ dimensional antisymmetric matrix M has an eigenvalue 
$\lambda_0$. Then we have, \be
{\bf \rm Det}[-{\bf 1} \lambda_0 - M] = (-)^n~{\bf \rm Det}[ {\bf 1} 
\lambda_0 - M]=0 \label{e1} 
\ee 
implying that $-\lambda_0$ is also an eigenvalue. Thus in this case we have 
two non-zero solutions of $\l^2_0$. Hence, four eigenvalues are grouped 
into two sets, $\{\n_e,\n_\tau\}$ and $\{\n_\mu,\n_s\}$ \footnote{At this 
stage we could have also chosen the pairs as $\{\n_e,\n_s\}$ and 
$\{\n_\tau,\n_\mu\}$, because they are indistinguishable from the point of 
view of the mass matrix as long as antisymmetry is unbroken. At later 
stage this choice will be justified by the mixing angles favoured by 
experiments.}. Each set having a pair of eigenvalues with equal magnitude 
and opposite sign as a result of the antisymmetry independent of the 
entries of the matrix! This guarantees a mass squared degeneracy among 
each set, hence solar neutrino problem can in principle be described by 
$\n_e \leftrightarrow \n_\tau$ oscillations and atmospheric neutrino 
problem by the $\n_\mu \leftrightarrow \n_s$ oscillations. Now we are in
a position to include Majorana mass terms to include see-saw 
mechanism\cite{seesaw} which explains the smallness of neutrino mass. We 
present a modified model to include Majorana terms and thereby see-saw 
mechanism. The skeleton key to the following discussion is that the 
eigenvalues of a matrix $MM^\dagger$ are the squares of those of $M$. The 
see-saw mechanism suppresses the Dirac mass term and we get back a 
light left handed Majorana neutrino mass matrix, 
\be
M^\n_{ij}= ({V^2_F \over 2~V_R })~({\tan^2 \beta 
\over 1 + \tan^2 \beta}) ~M^D_{ik} ~[M^M]^{-1}_{kl} 
{M^D}^{\dagger}_{lj} \label{e6} 
\ee
Where $\tan \beta \equiv {<H_u> \over <H_d>}$. $M^D$ and $M^M$ are the 
Dirac and right handed Majorana type Yukawa texture. It is understood in 
the above formula that we are working in a left-right symmetric theory or 
a grand unified theory as SO(10) which embeds left-right symmetric gauge 
group. In such a case $V_R$ is the scale in which the right handed 
symmetry is broken. In a supersymmetric SO(10) model the right handed 
symmetry breaking scale is closely tied from the requirements of gauge 
coupling unification\cite{rgelr}. We note that if $M^M$ 
is an approximately diagonal matrix, the light neutrino mass eigenvalues 
keep the underlying pattern dictated by those of the Dirac mass 
textures $\epsilon$. Consequently we postulate a $4 \times 4$ Dirac and 
Majorana mass textures as, \bea
&& M^D=\pmatrix{0 & 1 & 1 & 1 \cr
                                  -1 & 0 & 1 & 1 \cr
                                  -1 & -1 & 0 & 1 \cr
                                  -1 +\epsilon& -1+\epsilon & -1+\epsilon 
& \ep } \nonumber \\ 
&& ~~M^M = 
\bordermatrix
{& \n^\tau_{R} & \n^e_R & \n^\mu_{R} & \n^s_R \cr
\hline 
&&&&\cr
\overline{({\n^\tau_{R}})^c}    & 1 & 0 & 0 & 0 \cr
 \overline{({\n^e_{R}})^c}    & 0 & 1 & 0 & 0 \cr
 \overline{({\n^\mu_{R}})^c}  & 0 & 0 & 1 & 0 \cr
 \overline{({\n^s_{R}})^c} & 0 & 0 & 0 & 1+ \eta \cr
}
\label{e} \eea 
Where $1+\eta$ is the mass of the sterile neutrino. We expect it to be 
a little different from the other right handed masses. Using the 
expression for the Dirac and Majorana masses given in Eqn(\ref{e}) and 
inserting them to Eqn(\ref{e6}) we can describe the $4 
\times 4$ light neutrino Majorana mass matrix in terms of five parameters 
$V_R$, $\tan~\beta$, $\eta$ $\epsilon$ and $\epsilon^\prime$. 
Note that the symmetry in the Majorana sector is broken by the parament 
$\eta$. While diagonalizing Eqn. (\ref{e6}) the breaking in the symmetry 
propagates to the $\nu_\tau-\nu_e$ sector as $\eta^2$. Hence if we choose 
$\eta \sim 
10^{-2.5}$ the $e-\tau$ sector has a tiny mass difference of $\eta^2 \sim 
10^{-5}$ as we desire to achieve. In the Dirac sector the fourth column and 
row of $M^D$ has to be generated from a singlet Higgs. We expect that the 
antisymmetry in the Dirac sector will be broken in the fourth row and 
fourth column. Thus we have parametrized the departure from antisymmetry 
by the parameters $\epsilon$ and $\epsilon^\prime$. 

To set the kernel of the following discussion let us consider a toy 
example of two generations. Let us say the mixing matrices of the neutrino 
and the charged leptons as,
\be
O^N=\pmatrix{\cos \theta_n & \sin \theta_n \cr
             -\sin \theta_n & \cos \theta_n}~~~O^L=\pmatrix{\cos \theta_l 
& -\sin \theta_l \cr \sin \theta_l & \cos \theta_l}
\ee
It is nice to observe that the combined mixing matrix $O=O^N 
{O^L}^\dagger$ can be expressed as,
\be
O=\pmatrix{\cos (\theta_n \pm \theta_l) & \sin (\theta_n \pm \theta_l) \cr
           -\sin (\theta_n \pm \theta_l) & \cos (\theta_n \pm \theta_l)}
\ee
such that the combined mixing angle $\theta= \theta_n \pm \theta_l$. The 
negative sign in $\theta_n \pm \theta_l$ is also relevant as we could have 
represented $O^L$ as ${O^L}^\dagger$ to begin with keeping all the 
physics unchanged. Note that if $\theta_u \sim \theta_d \sim \pi/4$ then 
$\sin 2 \theta \sim 0$ and in case either $\theta_n$ or $\theta_l$ 
vanishes we will have $\sin 2 
\theta \sim 1$. These are precisely the two regions of mixing angles we 
are seeking for the neutrino sector. For solar neutrino as well as LSND 
observations we need a tiny mixing angle compared to the case of 
atmospheric neutrino case where the mixing has to be maximal. Furthermore 
if the atmospheric neutrino oscillation is caused by the transition 
$\nu_\mu \leftrightarrow \nu_s$ transitions we have an added advantage. 
The charged leptonic partner of the sterile neutrino is absent from the 
model by construction. Consequently, if we can set the angle in the 
neutrino sector to be $\pi/4$ it will survive after the multiplication 
with the rotation matrix  of the charged leptons. On the other hand for 
the other transitions the total mixing angle will be $0$ or $\pi/2$ 
leading to vanishing $\sin 2 \theta$. In the realistic case of four fermion 
mixing, this effect can be less transparent. In any case we will perform 
a numerical study of the mixing angles. However, if we factorize the 
complete mixing matrix in rotations along mutually orthogonal planes the 
`interference' of neutrino rotations with those of the charged leptons 
may still be visible to some extent. A $4 \times 4$ unitary matrix can be 
factorized in terms of six mixing (rotation) angles and two phases. Let 
is, for simplicity, set the phases to be zero. If we denote $\sin 
\theta_{ij}$ and $\cos \theta_{ij}$ as $s_{ij}$ and $c_{ij}$ 
respectively, a possible factorization\cite{barger} is,

{ \small \begin{eqnarray}
O&=&\pmatrix{1 & 0 & 0 & 0\cr
           0 & 1 & 0 & 0 \cr
           0 & 0 & c^1_{34} & s^1_{34} \cr
           0 & 0 & -s^1_{34} & c^1_{34}}
  \pmatrix{1 & 0 & 0 & 0 \cr
           0 & c^1_{23} & s^1_{23} & 0 \cr
           0 & -s^1_{23} & c^1_{23} & 0 \cr
           0 & 0 & 0 & 1}
  \pmatrix{c^1_{12} & s^1_{12} &0 &0 \cr
           -s^1_{12} & c^1_{12}&0 &0 \cr
           0 & 0 & 1 & 0 \cr
           0 & 0 & 0 & 1 } \nonumber\\
&&\pmatrix{1 & 0 & 0 & 0\cr
           0 & 1 & 0 & 0 \cr
           0 & 0 & c^2_{34} & s^2_{34} \cr
           0 & 0 & -s^2_{34} & c^2_{34}}
  \pmatrix{1 & 0 & 0 & 0 \cr
           0 & c^2_{23} & s^2_{23} & 0 \cr
           0 & -s^2_{23} & c^2_{23} & 0 \cr
           0 & 0 & 0 & 1}
\pmatrix{1 & 0 & 0 & 0\cr
           0 & 1 & 0 & 0 \cr
           0 & 0 & c^3_{34} & s^3_{34} \cr
           0 & 0 & -s^3_{34} & c^3_{34}}
\end{eqnarray}
}
Let us interpret the $ 4 \times 4$ leptonic rotation matrix O in the basis, 
where the top most row stands for the $\tau$ generation second row stands 
for the $e$ generation third row stands for the $\mu$ generation and the
last row represents the sterile neutrino. It is interesting to note 
(which is sparsely appreciated in the  recent literature on 4 neutrino 
oscillations) that in this representation there can be two different 
angles, with superscripts 1 and 2, which are relevant for the LSND 
oscillations and three different angle, with superscripts 1, 2 and 3 
which are relevant for the atmospheric neutrino oscillations. Let us 
focus on the atmospheric neutrinos for a moment. The muon neutrinos, 
produced in the gauge basis, can take up any three rotation angles to 
oscillate into the sterile neutrino and thus escape detection.
  
Let us return to the factorization of the rotation matrix $O^L$ given in 
Eqn. (\ref{ol}) after suitable reordering of the rows, and that of $O^N$ 
which can be gotten by numerically diagonalizing the matrix $M^\nu$ given 
in Eqn. (\ref{e6}) and compare with the factorization of $O=O^N 
{O^L}^\dagger$. Below we quote the factorization of $O^L$ in the first 
line (see also Ref\cite{fritzch}) $O^N$ in the second line (as already 
discussed in Ref\cite{anti} to some 
extent but we have changed the parameter a little bit) and the combined 
mixing angles of $O$, which is a main result of this paper. In the 
following we have denoted $S^k_{ij} \equiv \sin^2~2\theta^k_{ij}$
\begin{eqnarray}
& {\rm Leptons} \rightarrow 
S^1_{12}=0.913~S^1_{23}=0.991~S^2_{34}=0.00 & \\
& {\rm Neutrino} \rightarrow 
S^1_{12}=0.891~S^1_{23}=0.738~S^2_{34}=0.996 & \\
&----------------------& \nonumber \\
& {\rm Combined} \rightarrow 
S^1_{12}=0.010~S^1_{23}=0.079~S^2_{34}=0.996 & \label{result}
\end{eqnarray}
We have suppressed $S^2_{23}$, $S^1_{34}$ and $S^3_{34}$. Our purpose is to 
show the existence of relelant mixing angles for solar$\equiv S^1_{12}$, 
LSND$\equiv S^1_{23}$ and atmospheric$\equiv S^2_{34}$ neutrino 
oscillation experiments. We have used the parameter space 
$\{\tan\beta=2;V_R=10^{14.1};\epsilon=\eta=10^{-2}; 
\epsilon^\prime=-10^{-3}\}$ and the mass squarred differences are given by
\be
\Delta~m^2_{e \tau}=10^{-5.06}; \Delta~m^2_{\mu s}=10^{-2.03}; 
\Delta~m^2_{e \mu}=0.91 ~~{\rm eV^2} \label{result1}
\ee
It is instructive to study the $4 \times 4$ neutrino mass matrix obtained 
from the formula of Eqn. (\ref{e6}) and using the parameters given above. 
Up to second decimal place, the light neutrino mass matrix is,
\be
 M^\n=\pmatrix{0.49 & 0.32 & 0 & 0.33 \cr
               0.33 & 0.49 & 0.33 & 0 \cr
               0 & 0.33 & 0.49 & 0.33 \cr
               0.33 & 0 & 0.33 & 0.49} {\rm eV}
\ee
The $<\n_e \n_e>$ Majorana mass term is the (3,3) element of the above 
matrix. It is barely at the upper limit given by the neutrino-less double 
beta decay experiments. Note that antisymmetry in the Dirac matrix has 
disappeared because of $M^\n$ is primarily $M^D {M^D}^\dagger$ however 
the eigenvalues keep the partial degeneracy as the eigenvalues of $M^\n$ 
are squares of those of $M^D$. The eigenvalues also scale inversely as 
$M_R$. Consequently a stricter upper bound given by neutrino-less double 
beta decay experiments will increase $M_R$ thus reducing the absolute 
value of the eigenvalues too. Thus a very strict upper bound on $<\n_e 
\n_e>$ Majorana mass will force the predictions out of range of that 
required by LSND measurements. A few more comments on the dark matter 
is in order. COBE measurements of the quardrupole anisotropy of cosmic 
microwave background can determine the fluctuation in CMB at the scales 
of roughly the present size of the universe. With this normalization of 
the Harrison-Zel'dovich power spectrum $\Omega_\nu=1$ HDM (reletivistic)
has too little power for galaxy formation and CDM (non-reletivistic) has 
much too power at small scales. A cold plus hot (CHDM) model thus fits 
the observations, when the neutrino mass is 5 eV. However a even more 
baroque scheme has been proposed with a pair of 2.5 eV neutrinos share 
the needed 5 eV. In this case the heavy pair $\nu_\nu-\nu_X$ can explain 
atmospheric neutrino anomaly much to the jubilation of our's. 
If we keep $\Omega_\nu$ fixed and distribute it over two mass-degenerate 
neutrinos, in practice, we get somewhat lower `power' on cluster scales 
(typically Mpc), which seems to fit the data somewhat better. 
Unfortunately, the neutrinoless double beta decay constraints can allow 
us a pair of neutrinos at most at 1 eV falling short of the desired mass. 
We await the calculations\cite{primack1} of $\Omega_\nu \approx 0.4, 
\Omega_\Lambda=0.6$ cosmology. 

To conclude we have taken the point of view that the Yukawa couplings can 
be either zero or 1. In other words nature has not produced the variety 
of masses through arbitrary strengths of Yukawa couplings but through the 
structure of the textures; of course assuming that quarks and leptons are 
really fundamental particles. The structure of the textures is such that 
given the number of generations and hence the dimension of the matrices the 
relevant `off-diagonal' mixing angles are maximal or $\theta=\pi/4$. In 
particular for neutrino mixings the charged current interactions (which 
are the only way to catch the the fermioninc mixing angles) force us to 
measure only $V^N {V^L}^\dagger$ and so we measure $\theta_n \pm 
\theta_l$ in neutrino oscillations. The trick of this paper is that by 
construction the there is no sterile charged lepton. Hence if we look 
only at the charged lepton mass matrix the corresponding fictitious 
mixing angles arizing from the charged lepton mass matrix vanishes or 
$\theta_l=0$. Hence, in particular $\theta_{\mu s} \sim \pi/4 \pm 0$ 
remains as $\pi/4$ leading to $\sin^2~2\theta_{\mu s} \sim 1$. On the 
contrary the mixing in the $e-\tau$ and $e-\mu$ sectors are 
$\theta \sim \pi/4 \pm \pi/4$ leading to 
$\sin^2~2\theta_{e \tau} \sim 0 $ 
and $\sin^2~2\theta_{e \mu} \sim 0 $.   
The main symmetries of the Yukawa sector are the following. The neutrino 
Dirac Yukawa couplings are antisymmetric to the leading order, the 
Majorana Yukawa coupling is diagonal in the leading order and the charged 
leptonic Yukawa couplings are democratic in the leading order. The mass 
patterns arizing from these textures are previously studied in the 
literature. The democratic $3\times 3$ Yukawa matrix gives hierarchical 
mass pattern as desired ($m_\tau >> m_\mu >> m_e$) and maximal mixing 
angles whereas the antisymmetric Yukawa couplings in the neutrino Dirac 
sector gives bi-degenerate mass pattern ($m_{\nu_s} \sim m_{\nu_\mu} >> 
m_{\nu_e} \sim m_{\nu_\tau}$) and maximal mixings. We have quoted values of 
the symmetry 
breaking parameters which gives rise to $\Delta m^2_{e\tau}=10^{-5.06}$ 
eV$^2$ and $\Delta m^2_{\mu s}=10^{-2.03}$ eV$^2$ and $\Delta 
m^2_{e\mu}=0.91$ eV$^2$ relevant to solar atmospheric and LSND neutrino 
oscillation experiments whereas the corresponding mixing angles are  
$\sin^2~2\theta_{e \tau} = 0.01 $ $\sin^2~2\theta_{e \mu} \sim 0.079 $ 
$\sin^2~2\theta_{\mu s} \sim 0.99$. It is true that the appeal of the 
mechanism that we have stated here depend crucially on the establishment 
of the LSND results which forces us to consider a sterile neutrino in 
addition to the three known neutrino species which does not have a 
counterpart in the charge leptonic sector. Secendly, even though this 
mechanism of generating large and small mixing angles that we have 
proposed in this paper is quite generic; these particular set of textures 
can be established or ruled out by stricter upper bound on electron neutrino 
Majorana mass term from neutrinoless double beta decay experiments in 
combination with LSND results.
\vskip 1cm
\noindent I acknowledge discussions with P. Premysler on properties 
of matrices. I thank M. Bastero-Gil, R. N. Mohapatra, J. C. Pati and Q. 
Shafi for comments.

\end{document}